*Ability of Cyanobacteria and Arthrobacter Species to Remove Gold Ions from Solution*

E.Gelagutashvili,  E.Ginturi,  A.Rcheulichvili,  K.Tsakadze,  N.Bagdavadze, N.Kuchava , M.Djandjalia

I. Javakhishvili Tbilisi State University
E. Andronikashvili Institute of Physics
0177, 6, Tamarashvili S.,
Tbilisi, Georgia

Abstract

The biosorption of Au(III) – *Spirulina platensis* and Au(III) – *Arthrobacter* species (*Arthrobacter globiformis* and *Arthrobacter oxidas*) were studied at simultaneous application of dialysis and atomic absorption analysis. Also biosorption of Au(III) - *Spirulina platensis* at various pH were discussed.

Biosorption constants for Au-cyanobacteris *Spirulina platensis* at different pH, and for *Arthrobacter oxidas* and *Arthrobacter globiformis* at pH=7.1 are :

1.  $K = 3.91 \times 10^{-4}$     (Au- *Arthrobacter oxidas 61B,* pH=7.1*)*
2.  $K = 14.17 \times 10^{-4}$     (Au- *Arthrobacter globiformis  151B,* pH=7.1*).*
3.  $K = 2.07 \times 10^{-4}$     (Au- *Spirulina platensis* , pH=7.1)
4.  $K = 4.87 \times 10^{-4}$     (Au- *Spirulina platensis* , pH=6.2)
5.  $K = 8.7 \times 10^{-4}$      (Au- *Spirulina platensis,* pH=8.4)

**Introduction**

The metal-uptake ability of microorganisms has been known for a long time. Interest in the development of metal removal by biosorption using microorganisms is shown in literature [1-4].  In[1], it is suggested that carboxyl groups of the cell wall alginate play an important role in the cobalt binding. Apparently, chelation was the main mechanism of cadmium cation sequestration by the algal biomass, while ion exchange was the main mechanism of the nickel cation sequestration. Lead binding mechanism to the algal biomass included a combination of ion exchange, chelation and reduction reactions accompanied by metallic lead precipitation on the cell walls [2]. The reduction mechanism was suggested for gold cations by *Sargassum biomass*, tannin was the cell wall component suggested to play a role in the reduction of gold due to its strong reducing ability[3].

Although thousands of micro-algal species have been identified, very few have been investigated for their biosorption potential. Cyanobacteria *Spirulina platensis* is widely used as healthy food due its content of protein, vitamin and active substances for immune system [4]. *Spirulina platensis* was shown to be accumulating toxic metal ions from its environment [5]. Further, it was also indicated that some groups of algal cell biomass were responsible for binding to various ions [6,7].

Some bacteria have evolved mechanisms of heavy metal detoxification, and some even use them for respiration. The recovery of gold using algae cells were investigated in [8-10], and microorganism-gold interaction was investigated in [11,12]. Algae have tremendous role in bioremediation of toxic and precious metals and their bioconversion to different non-toxic forms [13]. They also can produce metal-nanoparticles [14], that have significant role in nanotechnology, DNA labeling and development of biosensors [15]. XRD analysis of gold nanoparticles confirmed the formation of metallic gold. Fourier transform infrared spectroscopic measurements revealed the protein is the possible biomolecule responsible for the reduction and capping the biosynthesized nanoparticle [14]. *Shawanella algae*, a Fe(III)-reducing bacterium, produces gold nanoparticles in anaerobic conditions[16].

In spite the fact that there is significant increase of the interest to study biosorption of metals by microorganisms, there is little information on the kind of microorganisms that have the ability of gold absorption.

In this paper, cyanobacteria *Spirulina platensis* and gram-positive bacteria *Arthrobacter species* were screened for their ability to adsorb gold from a solution containing hydrogen tetrachliroaurate (III). The effect of pH on gold biosorption by *Spirulina platensis* cells and gold-*Arthrobacter species*, which adsorbed significant amounts of gold from a solution containing hydrogen tetrachliroaurate (III) were studied by the methods of dialysis and atomic absorption analysis.

## Materials and Methods

The other reagents and bacteria were used: $HAuCl_4$ (Analytical grade), cyanobacteria *Spirulina platensis* and *Arthrobacter* bacterials *Arthrobacter oxidas 61B* and *Arthrobacter globiformis 151B*.

*Spirulina platensis* IPPAS B-256 strain was cultivated in a standard Zaroukh alkaline water-salt medium at a temperature of +34℃, illumination ∼ 5000 lux, initial pH 8.7 and at constant mixing.

*Arthrobacter* bacteria were cultivated in the nutrient medium [17].
Cells were centrifuged at 12000 rpm for 10 min and washed three times with phosphate buffer (pH 7.1). The centrifuged cells were dried without the supernatant solution until constant weight. After solidification (dehydration) of cells (dry weight) solutions for dialysis were prepared by dissolving in phosphate buffer. Known quantity of dried bacterium suspension was contacted with solution containing known concentration of the metal ions. For biosorption studies, the dry cell weight was kept constant (1 mg/ml). All experiments were carried out at $23^0C$ temperature.

The experiments of dialysis were carried out in 5ml cylindrical vessels made of organic glass. A cellophane membrane of 30μm width (type -"Visking" manufacturer - "serva") was used as a partition. The duration of dialysis was 72 hours. *Spirulina platensis* was used in suspension, dissolved in medium with pH=8.43, dissolved in phosphate buffer, pH=7.1, and in water pH =6.2. In addition, biosorption of Au-*Arthrobacter globiformis 151B and Arthrobacter oxidas* 61B dissolved in phosphate buffer pH=7.1 were also investigated.

In all mentioned cases, the concentration of *Spirulina platensis* was 400mg/100ml. The metal concentration ranged from $10^{-2}$M to $10^{-5}$M (hydrogen tetrachliroaurate (III) ($HAuCl_4$)). The metal concentration after the dialysis was measured by atomic-emission analysis at λ=242.8 nm wavelength. Each value was determined as an average of three independent estimated values ± the standard deviation. The processing of experimental data was made using Freundlich [18] model. $C_b=KC_{total}^{1/n}$, where $C_b$ is concentration of metal ion adsorbed, $C_{total}$ is equilibrium concentration of metal ion in solution. *K* and *n* are empirical constants, which are indicators of adsorption capacity and adsorption intensity and they have been calculated from intercept and slope of the plots. Freundlich equation can be linearized in logarithmic form and empirical constants can be determined.

## Results and Discussion

Fig. 1-2 shows the biosorption isotherms by using the fitted linearized Freundlich model. Fig.1 represents Freundlich isotherms for Au-*Spirulina platensis* at different pH=6.2 (6.2, 7.1, and 8. 4)) and fig.2 represents Freundlich isotherms for Au *Arthrobacter globiformis 151B and Arthrobacter oxidas 61B*. By means of Freundlich isotherms the biosorption constants (*K*) and the capacity (*n*) were determined for Au_*Spirulina platensis* and _*Arthrobacter species*. The biosorption constants and capacity values in addition to R and Standart deviation values are shown in table 1.

The results showed that, the ability of the microbial cells to adsorb gold ions significantly varied. It is clear from table 1, that biosorption of Au-*Spirulina platensis* depends on pH. In particular, the highest value of biosorption constant was observed at pH=8.4. pH has also an important effect on gold biosorption capacity. The effect of the initial pH could be related to different protonation of the functional groups present in the cell membrane. Dependence of Au sorption by *Spirulina platensis* on the medium pH can be attributed to their differences in chemical composition of cell surface leading also to variable chemical interactions [13]. The adsorption of heavy metals on the surface is usually attributed to the formation of complexes between metals and the carboxyl, amino, sulfhydryl funcrional groups. In[6], it was shown, that carboxyl groups on algal cell biomass are active sites for binding to various ions.

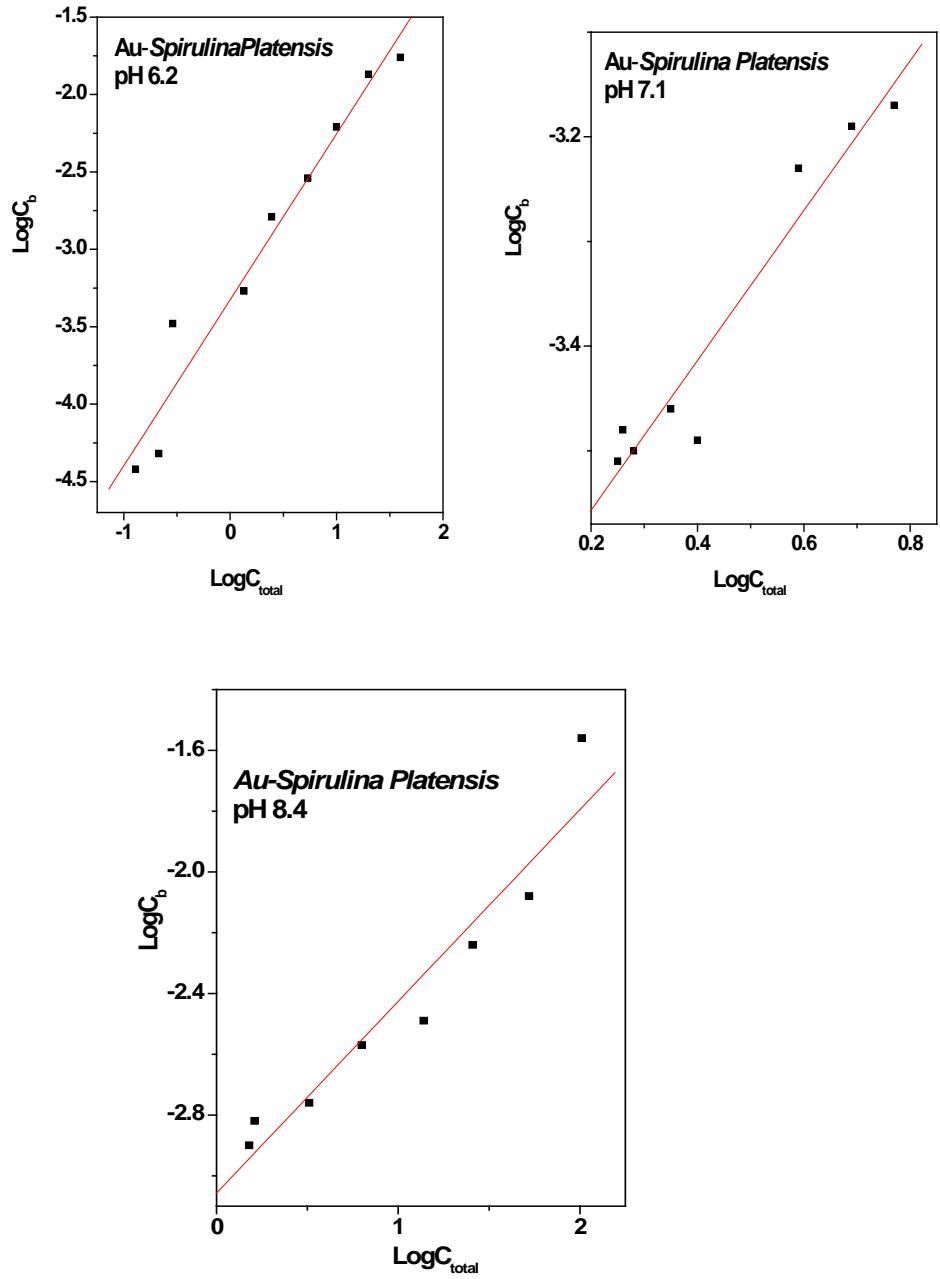

Fig.1. Freundlich isotherms for *Au-Spirulina platensis* at various *pH* .

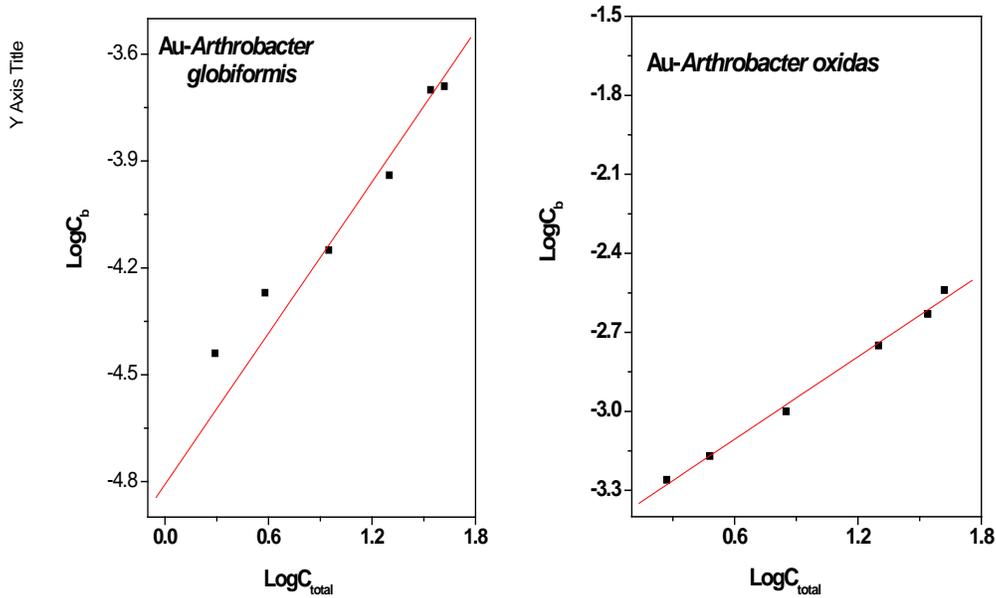

. Fig.2.   Freundlich isotherms for Au-*Arthrobacter species*    pH=7.1

Table 1. Biosorption parameters for gold by *Spirulina platensis* and *Arthrobacter species*

|  | Biosorption constant | Biosorption capacity | Standard deviation | Correlation coefficient |
|---|---|---|---|---|
|  | $K$ | $n$ | $SD$ | $R$ |
| *S. platensis* dissolved in medium $pH=8.4$ | $8.71 \times 10^{-4}$ | 1.59 | 0.13 | 0.96 |
| *S. platensis* dissolved in phosphate buffer    $pH=7.1$ | $2.07 \times 10^{-4}$ | 1.39 | 0.04 | 0.97 |
| *S.platensis* dissolved in water $pH=6.2$ | $4.87 \times 10^{-4}$ | 0.93 | 0.21 | 0.98 |
| *Arthrobacter globiformis*  151 B,   pH 7.1 | $14.17 \times 10^{-4}$ | 0.81 | 0.09 | 0.96 |
| *Arthrobacter   oxidas 61B, pH 7.1* | $3.90 \times 10^{-4}$ | 1.92 | 0.22 | 0.98 |

From table 1 it is clear also that there is   significant difference between the binding constants for Au –*Arthrobacter oxidas*  and Au-A*rthrobacter  globiformis*. Comparative Freundlich biosorption characteristics for Au *Arthrobacter* species show that the biosorption constant for *Arthrobacter globiformis* is higher, than that for *Arthrobacter oxidas.* On the contrary, capacity (*n*) for *Arthrobacter oxidas* is more, than that for A*rthrobacter globiformis*.  Correlation coefficient in all discussed cases is more, then 0.95.

*Arthrobacter globiformis* amine oxidase in the holo and apo forms adsorbed onto a Au (III) surface have been observed by scanning tunnelling microscopy (STM) at $23^0$C [19]. Individual protein molecules denaturate as they adsorb onto a bare Au surface, although they keep a dual appearance. STM voltage affects the distance between the units of denaturated proteins: negative voltages separate them while positive brings them together.

Comparative analysis between cyanobacteria and *Arthrobacter* species show that ability of *Arthrobacter globiformis* to remove gold from solution is better than that of *Spirulina platensis* at various pH and *Arthrobacter oxidas.*

The cell wall consists of variety of polysaccharides and proteins and hence offers a number of active sites capable to bind metal ions. Difference in the cell wall composition of different groups of microorganisms causes significant difference in the type and amount of metal ion binding to them.

This work was supported by grant STCU #4744